\documentclass[twocolumn,prb]{revtex4-2}
\pdfoutput=1
\usepackage{hyperref}
\usepackage{amssymb}
\usepackage{graphicx}
\usepackage{amsmath}
\usepackage{float}
\usepackage{epstopdf}
\usepackage{times}
\usepackage{color}
\usepackage{subfigure}
\usepackage{tensor}
\usepackage{bm}
\usepackage{slashed}

\newcommand{\Tr}{\text{Tr}}
\newcommand{\tr}{\text{tr}}
\newcommand{\beq}{\begin{equation}\begin{aligned}}
\newcommand{\eeq}{\end{aligned}\end{equation}}

\graphicspath{{./}{./figs/}}

\begin{document}

\title{Destabilization of U(1) Dirac spin-liquids 
on two dimensional non-bipartite lattices
by quenched disorder}


\author{Santanu Dey}
\affiliation{Institut f\"ur Theoretische Physik 
and W\"urzburg-Dresden Cluster of Excellence 
ct.qmat, Technische Universit\"at Dresden,
01062 Dresden, Germany}

\begin{abstract}
    The stability of the Dirac spin-liquid on 
    two-dimensional 
    lattices has long been debated. 
    It was recently demonstrated 
    [Nature Commun. {\bf 10}, 4254 (2019) and 
    Phys. Rev. B {\bf 93}, 144411 (2016)]
    that the staggered $\pi$-flux Dirac spin-liquid
    phase on the non-bipartite triangular 
    lattice may be stable in the clean limit. 
    However, quenched disorder plays a crucial
    role in determining whether such a phase is experimentally 
    viable. For SU(2) spin systems, the effective zero-temperature, 
    low-energy
    description of Dirac spin-liquids in 
    $(2+1)$ dimensions is given by the compact 
    quantum electrodynamics ($\rm cQED_{2+1}$) which admits monopoles. 
    It is already known that generic quenched 
    random perturbations to the non-compact version of 
    $\rm QED_{2+1}$ (where monopoles are absent) lead  
    to strong-coupling instabilities. In this paper we study
    $\rm cQED_{2+1}$ in the presence of a class of
    time-reversal invariant quenched disorder perturbations.
    We show that in this model, 
    random non-Abelian vector potentials make 
    the symmetry-allowed monopole operators more relevant.
    The disorder-induced underscreening of monopoles, thus, 
    generically makes the gapless spin-liquid phase fragile.   
\end{abstract}

\date{\today}

\maketitle

\section{Introduction}

Quantum spin-liquids with their topologically ordered 
ground states, fractionalized excitations, and 
long-range entanglements offer a fascinating
insight into many-body quantum
correlations \cite{wen02,lee08,balents10}. Experimentally, the
observation of a spin-liquid phase has been fraught with 
the complications arising from spatial inhomogeneities in 
real materials, which often leads to symmetry breaking towards a 
spin-glass ground state \cite{nakamura97,kimchi18,ma18}. Although 
the role of quenched disorder in a frustrated spin system may 
vary considerably \cite{knolle19,choi19,savary17,dey20}, in a number 
of examples \cite{tsomokos11,sen15,dey20b} it has been shown 
that the topological properties of frustrated systems are
considerably affected by quenched disorder. In this paper we 
consider the gapless Dirac spin-liquid state
with $2N$ flavors of matter fermions and compact
U(1) 
gauge symmetry and investigate its stability in the 
presence of random gauge fluctuations. 

As a prototypical
spin-liquid state with linearly dispersing
gapless fractionalized spinons and
minimally coupled 
compact
U(1) gauge fields, the Dirac spin-liquid 
state has been discussed 
as a parent state for different 
competing orders \cite{lee06,hermele05}, 
deconfined quantum critical points between topological
phases \cite{grover13}, and
as the prospective ground states of the kagome lattice
Heisenberg model
\cite{ran07,hermele08} and the triangular
lattice Heisenberg model with next-nearest-neighbor
exchange interaction \cite{iqbal16}.
The variational Dirac spin-liquid state can be derived
from the mean-field decomposition of the
SU(2) Heisenberg Hamiltonian,
\beq 
H=
\sum_{i,j}J_{ij}\vec{S}_i
\cdot\vec{S}_j
\eeq 
in terms of fermionic spinons. Here, $J_{ij}$ are exchange couplings 
between nearest- and next-nearest-neighbor spins.    
In this picture, a 
spin-1/2 operator at site $i$ is rewritten as
$\vec{S}_i=(1/2)f^\dagger_{i,\alpha}\vec{\sigma}_{\alpha\beta}f_{i,\beta}$
with the physical constraint $
\sum_{\alpha}f^\dagger_{i,\alpha}f_{i,\alpha}=1$.
Here $f_{i,\alpha}$ are fractionalized fermionic spinons with 
$\alpha = \uparrow,\downarrow$ 
being
the spin indices. The mean-field
decomposition with with bond variables
$t_{ij}=-\langle f^\dagger_{i,\alpha}f_{j,\alpha}\rangle$
reduces the Hamiltonian to the quadratic form 
\beq 
H_{\rm MF}=\sum_{ i,j}
\frac{J_{ij}}{2}
\left[|t_{ij}|^2
+
\left(t_{ij}
f^\dagger_{i,\alpha}f_{j,\alpha}+\rm H.c.\right)
\right]
,
\eeq 
with the
mean-field ansatz of bond variables $t_{ij}$ chosen suitably to 
minimize the variational energy. In 
the spinon decomposition, the compact U(1) gauge symmetry
is manifest with the transformation 
$f_{i,\alpha}\rightarrow e^{iA_i}f_{i,\alpha}$, which
ultimately leads to the emergence of
dynamical U(1) gauge field fluctuations. 
The emergent gauge group is necessarily compact as it is a subgroup 
of the larger compact SU(2) gauge group
related to the fractionalization of the physical SU(2) spins \cite{polyakov77,wen02}. 

On honeycomb, kagome and triangular lattices, a Dirac dispersion band 
of the spinons are realized with a suitable choice of nearest-neighbor 
mean-field parameter $t_{ij}$. However, the fluctuations around the 
mean-field state may destablize it and in that case the mean-field 
spin-liquid state does not correspond to a physical state of the 
original spin Hamiltonian. In the triangular lattice the
nearest-neighbor $\pi$-flux 
mean-field ansatz with no fluxes
through the lower triangular plaquettes, 
$\prod_{(ij)\in \triangledown}t_{ij}=1$ and $\pi$
fluxes through the upper triangular plaquettes 
$\prod_{(ij)\in \triangle}t_{ij}=-1$ yield a Dirac spin-liquid state
with four gapless Dirac cones, i.e., two Dirac nodes 
(valleys) for each spin flavor. Variational Monte Carlo 
studies \cite{iqbal16,kaneko14} have indicated that on the 
$J_1\mbox{-}J_2$ next-nearest neighbor triangular 
lattice Heisenberg antiferromagnet, the Dirac spin-liquid state is  
energetically favorable compared to other magnetically ordered and 
spin-liquid type states across a certain region of the parameter space.
As a clear demonstration of the 
stability (against fluctuations) and energetic favorability of the
U(1) Dirac spin-liquid awaits for other lattices, we presently focus 
on 
the microscopic realization of the Dirac spin-liquid ground state of the 
triangular lattice 
and 
consider quenched random perturbation to its clean limit. However, as our 
disorder 
study is performed in the continuum, the findings are applicable to all 
Dirac spin-liquid ground states with compact U(1) gauge field fluctuations. 

At zero temperature and in the long-wavelength low-energy limit, 
the 
Dirac spin-liquid state and its gauge fluctuations are described 
by the action of the $(2+1)$ dimensional compact quantum 
electrodynamics \cite{herbut02,hermele05},
\begin{equation}
\begin{aligned}
S_{\rm cQED} &= \int d\tau\ d^2r\
\left[
\bar{\psi_i}\gamma^\mu
\left(\partial_\mu+i A_\mu\right)
\psi_i
+\frac{1}{4e^2}
{F_{\mu\nu}}^2
\right],
\end{aligned}\label{eq:qed3}
\end{equation}
where $\psi$ are $2N$ copies of two-component fermionic fields
which are descendants of the fermionic spinors $f_{i,\alpha}$
with $i\in 1,\dots,2N$ and 
$F_{\mu\nu}=\partial_\mu A_\nu-\partial_\nu A_\mu
$ is the usual field strength tensor for a Maxwell 
gauge theory. The number $N$ is determined by the
number of Dirac nodes of the microscopic dispersion
per spin, i.e., 
$N=2$ for the triangular lattice Dirac spin-liquid.
In the following, we will suppress
all the fermionic flavor indices. 
Here the three Dirac $\gamma$ matrices
$\gamma^\mu$ are taken to be two-component \cite{kim99,song19} and
they obey the usual Clifford algebra,
$\{\gamma^\mu,\gamma^\nu\}=2\delta^{\mu\nu}I_2$.
The gauge charge has 
scaling dimension $\left[e^2\right]=+1$ and it
flows to infinity in the deep infrared. 
Consequently, the infrared fixed point of the action 
as written 
has conformal symmetry in the large-$N$ limit \cite{hermele04}. The
action also has an emergent SU($2N$) symmetry under which the
fermions $\psi$ transform as vectors.

From the lattice regularization, the gauge fields 
$A_\mu$ are $2\pi$ periodic compact variables and 
therefore it can be shown that the  
$\rm cQED_{2+1}$
action must admit
monopole operators of charge $q$ which 
insert $4\pi q$ units of magnetic flux locally
\cite{polyakov77,nogueira08,borokhov02}.
It was shown originally by Ref.~\onlinecite{polyakov77} 
that the proliferation of these monopoles strongly confines 
the electric charges of the pure compact Maxwell gauge theory. The 
argument follows from considering a dilute gas of elementary 
monopoles $q=\pm 1/2$ which can be described by the three-dimensional 
sine-Gordon model in the continuum limit \cite{nogueira08,hermele04},
\beq
S_{\rm sG}=\int d^3r
\left[\frac{1}{2}
\left(\frac{e}{2\pi}\right)^2
\left(\partial_\mu\chi\right)^2
-2y\cos \chi
\right].\label{eq:monopole_gas}
\eeq 
In the Hamiltonian picture, the operator $e^{i\chi}$ adds $2\pi$ 
magnetic flux and creates a monopole operator in three space-time 
dimensions. Thus, following the Coulomb gas picture $y$ is interpreted 
as the the monopole fugacity. In the absence of any matter coupling 
the monopole fugacity with the scaling dimension, $\left[y\right]=3$ 
flows to strong coupling in the infrared and confines the pure gauge 
theory. However, in the presence of matter fields coupling the 
Coulomb interaction between the monopole charges are screened 
with a modified renormalization of the monopole fugacity \cite{hermele04},
\beq
\frac{dy}{dl}=(3-\Delta_{\mathcal M})y.
\label{eq:fugacity_rg}
\eeq
Here, $\Delta_{\mathcal M}$ is the effective scaling dimension of the 
monopole creation operator $e^{i\chi}$ in the presence of the matter 
coupling. The relevance of the monopole fugacity operator now depends on
whether $\Delta_{\mathcal M}$ is less than space-time dimension $3$. 

The fate of the compact U(1) gauge theories minimally coupled to
gapless fermion spinons with Dirac dispersion 
has been controversial \cite{herbut03,ioffe89}. However, it has been shown 
the (anomalous) scaling dimension of the monopole operator in the 
presence of a large (even) number of $2N$ fermionic spinons grows as 
$\Delta_{\mathcal M}\propto 2N$, indicating a stable 
deconfined phase sufficiently large fermionic flavors
\footnote{However, more generically, the gauge charge 
$e^2$ may flow to a fixed point instead of the strong coupling in the 
presence of matter fields, yielding a critical conformal phase 
\cite{hermele04}.}
\cite{hermele04,nogueira08}.In particular, within a large $2N$ 
approximation the scaling dimension of the monopole operators is 
found to be of the form \cite{dyer13}
\begin{equation}
    \begin{aligned}
        \Delta_{\mathcal{M}^{(q)}}
        =(2N)\lambda_0^{(q)}+\lambda_1^{(q)}+O(1/2N).
    \end{aligned}
\end{equation}
For the lowest charge $q=1/2$ monopole operators, computation using 
state-operator correspondence \cite{borokhov02,pufu14}
have yielded, 
$\lambda_0^{(1/2)}=0.265$ and $\lambda_1^{(1/2)}=-0.0383$.
Therefore, if the monopole operators of the lowest charge are 
allowed, the minimum number of fermionic flavors needed to avoid 
confinement is $2N_C\geq 12$. This number is more than the number of 
fermionic flavors obtained in the 
known
mean-field Dirac spin-liquid states
($2N=4$ for the kagome \cite{hermele08} and triangular lattice \cite{iqbal16}
; and $2N=4$ and $2N=8$, 
respectively, for the staggered and $\pi$ flux 
Dirac spin-liquid
states in square lattices \cite{marston89,hermele05}). 
However, for 
the Dirac spin-liquid states in nonbipartite triangular and kagome lattice 
geometries, it has been recently shown that the monopole operators of the 
lowest charges are prohibited by lattice symmetries \cite{song19,song20}. 
Ref.~\onlinecite{song19} demonstrated that for the triangular lattice
Dirac spin-liquid only monopole operators with charges $q\geq 3/2$ are
allowed by microscopic symmetries and these higher charge
monopole operators are all irrelevant if the large 
$2N$ approximated monopole scaling dimension $\Delta_{\mathcal{M}^{(q)}}$
\cite{pufu14}
is extrapolated to $2N=4$. 
This indicates the possibility of a stable deconfined Dirac spin-liquid phase 
in the triangular lattice. 
The same analysis found that for the kagome lattice the smallest allowed 
monopole operators are very close to being marginal but relevant
within the large $2N$ 
approximation. Indeed, on the triangular 
lattice next-nearest-neighbor
$J_1\mbox{-}J_2$ Heisenberg model the Dirac spin-liquid phase is found to 
be stable and energetically favorable in variational
Monte Carlo simulations \cite{kaneko14,iqbal16}
and density matrix renormalization (DMRG) calculations
\cite{hu19}. A spin-liquid phase
was found to be stable for $0.07<J_2/J_1<0.15$
by a separate 
DMRG study \cite{zhu15}. In other reports \cite{hu15,wietek17}, 
a chiral spin-liquid is found in
the same parameter range in the presence of a time-reversal
symmetry-breaking perturbation, which is consistent
with a viable Dirac spin-liquid phase in the time-reversal symmetric 
limit.

In our treatment, we examine the fate of the 
${\rm cQED}_{2+1}$ 
action [Eq.~\eqref{eq:qed3}] as an effective theory of the 
Dirac spin-liquid
in the presence of time-reversal symmetric microscopic
perturbations. Microscopically, the 
triangular lattice Dirac spin-liquid
will be our focus as a promising candidate in the 
clean limit.
Theoretical efforts \cite{thomson17,goswami17,zhao17} to study
${\rm QED}_{2+1}$ in the presence of various quenched random 
perturbation has so far been focused
on the non-compact limit which neglects
the monopole operators. 
It has been established that random perturbations
which break the time reversal symmetry and/or break completely
the emergent 
SU($2N$) symmetry 
of the $\rm\ cQED_{2+1}$ action
drive a renormalization group (RG) flow to a strong disorder 
coupling fixed point, which in the microscopic sense
indicates the destruction
of the spin-liquid \cite{thomson17}.
In this paper, we therefore focus on time-reversal symmetric disorder
which breaks the SU($2N$) symmetry only partially. It has been shown
that weak random perturbations 
which break the SU($2N$) symmetry down to U(1)$\times$SU($N$)
flow to a finite disorder conformal fixed line 
\cite{thomson17}
and, consequently, the Dirac spin-liquid phase
may be expected to survive \cite{thomson17,vafek08}. 
In this context,
we consider the
compact nature of the effective theory and
perturbatively calculate the
disorder induced modification to the scaling dimension of the
monopole operator to further clarify
the fate of the algebraic Dirac spin-liquid phase. 

In Sec.~\ref{sec:disorder_model} we 
introduce the RG marginal 
random couplings that 
we consider as a perturbation to the 
$e^2\rightarrow\infty,
N\rightarrow\infty
$ conformal fixed point of the theory and discuss their microscopic
origin. 
Adapting the state-operator correspondence method described in
Sec.~\ref{clean_sdim}, in Sec.~\ref{dirty_sdim} we calculate
the  scaling dimension of the monopole operators in the dirtied 
${\rm\ cQED}_{2+1}$ within a controlled expansion in large-$N$
and perturbative disorder strength. We find that disorder significantly
reduces the scaling dimension of the monopole operators and 
enhances the possibility of confinement of the spinons which carry
electric gauge charges. In Sec.~\ref{rg_flow} we consider the 
combined flow of the monopole fugacity and the perturbative disorder 
couplings and show that even when disorder in itself
remains marginal, the monopole fugacity may flow to strong coupling 
and confine the theory. In the concluding Sec.~\ref{conclusion}, 
we comment on the 
instabilities introduced by disorder-driven spinon confinement within
the context of the Dirac spin-liquid phase and argue why among
other possibilities a glassy random-singlet like
ground state is a likely outcome for even small to moderate disorder 
in this scenario.

\section{Quenched disorder in ${\rm cQED}_{2+1}$}
\label{sec:disorder_model}

The ${\rm\ QED}_{2+1}$ action is an effective low-energy description
and the spatial inhomogeneities in the lattice translate to
random coupling perturbations to the theory. Ref.~\onlinecite{thomson17} 
showed that there are no relevant 
random
perturbations to ${\rm\ QED}_{2+1}$ in 
the large $(2N)$ limit and the only marginal 
random
couplings are the various 
conserved currents and mass operators associated with the SU($2N$) 
symmetry of the fermions \cite{thomson17}. In our discussion, we  
choose $\sigma^\alpha$ and $\tau^b$ as the $(2N)^2-1$ generators 
of $\rm SU(2N)$ where
$\sigma^\alpha$ are $2\times 2$ Pauli matrices 
with $\alpha=x,y,z$
and 
$\tau^b$ are $N^2-1$ traceless $N\times N$ 
Hermitian matrices with the normalization 
$\tr\left[\tau^\alpha\tau^\beta\right]=\delta^{\alpha\beta}
/2$. The generators satisfy the usual commutation relations
$[\sigma^\alpha,\sigma^\beta]=i\epsilon_{\alpha\beta\gamma}\sigma^\gamma$
and $[\tau^a,\tau^b]=i\tensor{f}{^{ab}_c}\tau^c$, where
$f_{abc}$ are structure constants of the corresponding SU($N$)
Lie algebra.
In this notation $\sigma^\alpha$ operate on the spin space and 
$\tau^\beta$ operate on the 
fermion-doubled
valley space originating
from the Dirac node structure of the parent mean-field state.
Associated with these symmetry generators are SU($2N$) current,
\begin{equation}
\begin{aligned}
&J^{\alpha b}_\mu=i\bar{\psi}\sigma^\alpha\tau^b\gamma_\mu\psi,
J^{\alpha 0}_\mu=i\bar{\psi}\sigma^\alpha\gamma_\mu\psi,
J^{0 b}_\mu=i\bar{\psi}\tau^b\gamma_\mu\psi,&
\end{aligned}
\end{equation}
and mass terms,
\begin{equation}
\begin{aligned}
&M^{\alpha b}=\bar{\psi}\sigma^\alpha\tau^b\psi,
M^{\alpha 0}=\bar{\psi}\sigma^\alpha\psi,
M^{0 b}=\bar{\psi}\tau^b\psi.&
\end{aligned}
\end{equation}
It is to be noted that terms not containing $\sigma^\alpha$ are
related to the 
spin-singlet local (bilinear) operators of the microscopic model
whereas the rest maps to the spin-triplet operators.
In the clean limit the conserved SU($2N$) currents, e.g.,
$i \bar{\psi} \sigma^\alpha\gamma^\mu \psi$ have the scaling
dimension $\Delta=2$ to all order in $1/(2N)$ but the SU($2N$)
mass terms e.g., $ i \bar{\psi} \sigma^\alpha \psi$ acquire
anomalous scaling dimensions 
$\propto 1/(2N)$ \cite{hermele05}. Let us
consider quenched random coupling to an arbitrary operator 
$O(\vec{r},\tau)$ such that the perturbing action
is 
$S_{\rm dis}=\int d\tau d^2r \ 
h(\vec{r})O(\vec{r},\tau)$ with uncorrelated 
random conjugate fields 
$\overline{h(\vec{r})h(\vec{r}')}=
\rho_{O}
\delta^{(2)}(\vec{r}-\vec{r}')$.
Following standard replica technique, 
$\overline{F}=\overline{\ln Z}=\lim_{n\rightarrow 0}
(Z^n-1)/n\sim \lim_{n,\rightarrow 0}
\prod_{r=1}^n Z_r$, a replicated partition summation emerges,
\begin{equation}
\begin{aligned}
Z_{\rm replica}
=&\int\mathcal{D}[\psi_r,A_r]
\exp
	\Big(-
	\sum_r
	\int d\tau d^2r\
\psi_r[\slashed{\partial}
+i\slashed{A_r}]\psi_r\\
&
+
\rho_{O}
\sum_{rs}
\int d\tau d\tau' d^2r\
O_r(\vec{r},\tau)
O_s(\vec{r},\tau)
\Big).
\end{aligned}
\end{equation}
From power-counting it clearly follows that 
$\Delta_{\rho_{O}}
=2+2z-2\Delta_{O}
$ where $z=-[\tau]$ is the dynamical 
critical
exponent. Therefore, in the 
large-$N$ limit when $z=1$, random couplings to 
the various SU($2N$) current and
mass terms are marginal at the tree level. Similarly, the
random couplings to
simple mass terms $\sim\bar{\psi}\psi$ are also RG marginal but
such mass terms break time-reversal symmetry in the $(2+1)$ dimension.
Quenched disorder breaks Lorentz invariance and, 
consequently, the scaling dimension of both 
the random SU($2N$)
mass and current
disorder 
couplings are modified beyond the tree level. 
In the absence of monopoles, previous works 
\cite{ludwig94,thomson17,zhao17,goswami17} have 
established that 
if such random couplings break the fermionic SU($2N$) symmetry or
the time-reversal symmetry the combined RG flow
generically
moves to a 
strong coupling fixed point. However, 
Ref.~\cite{thomson17} has shown that 
for time-reversal symmetric random perturbations,
if the symmetry is only
partially broken to U(1)$\times$ SU($N$), a 
finite disorder conformal fixed line is obtained,
parametrized by the corresponding
coupling strengths.
Technically, this fixed line is demarcated by the breakdown of the 
microscopic SU(2) symmetry down to U(1). 

In keeping with the goal of 
calculating the scaling dimension of the monopole operators by 
invoking the state-operator correspondence of radial quantization 
\cite{rychkov16}, we will presently only consider the 
SU($N$) symmetric, random current (RC) perturbations,
\begin{equation}
    \begin{aligned}
        S_{\rm dis}
        &=\int d\tau d^2 r\ 
        V_{\alpha j}(\vec{r})\
        i\bar{\psi}
        \sigma^\alpha\gamma^j
        \psi(\vec{r},\tau),\\
        P[V]
        &=\exp\left[-\frac{1}{2\rho_\alpha}
        \int d^2r\
        V_{\alpha j}^2(\vec{r})\right],
    \end{aligned}\label{eq:dis_pert}
\end{equation}
where a Gaussian distribution for the conjugate random field 
$V_{\alpha j}$ has been considered for convenience 
with $\rho_\alpha$ being the corresponding disorder strength. 
The index $j$ of Dirac matrices 
here runs strictly over the spatial components as the disorder is static 
in space. 

Microscopically, the time-reversal invariant local random 
perturbations are usually 
either random bond type, $P_{ij}=\vec{S}_i\cdot\vec{S}_j$
or vector-chirality type $\vec{C}_{ij}=\vec{S}_i\times\vec{S}_j$.
With time-reversal invariance,
the former behaves as scalars in spin space and, therefore, is
associated with the spin-singlet mass and
current terms $J_j^{0b}$, 
$M^{0b}$, and random Abelian vector potentials, whereas the latter
is associated with the spin-triplet mass and current terms 
$J^{\alpha 0}_j$ and $M^{\alpha 0}$. 
Although the random Abelian vector potential is 
known to be  an
irrelevant perturbation for non-compact $\rm\ QED_{2+1}$
\cite{thomson17,zhao17}, 
Ref.~\onlinecite{thomson17} 
showed that the random spin-singlet
SU($2N$)
current and mass terms are, 
however, relevant perturbations, and, therefore, it can
be surmised that random bond like perturbations are destructive to
the Dirac spin-liquid phase. On the other hand, the same treatment revealed 
the presence of a fixed line for 
U(1)$\times$SU($N$) symmetric 
random couplings to the spin-triplet 
terms $J^{\alpha 0}_j$ and $M^{\alpha 0}$, the former of which we
presently consider. In $(2+1)$ dimensions random the random 
current terms,
$J^{\alpha b}_j=i\bar{\psi}\sigma^\alpha\tau^b\gamma_j\psi$
preserve the time reversal symmetry \cite{thomson17}. It is to be 
noted that we are only considering time-reversal symmetric, static 
disorder in this treatment, the origin of which lies in nonmagnetic 
structural impurities. 

\section{Monopole scaling dimension of clean ${\rm cQED}_{2+1}$}
\label{clean_sdim}

In the absence of monopole operators,
the $\rm\ cQED_{2+1}$ action [Eq.~\eqref{eq:qed3}] has an additional 
topological symmetry $\rm U(1)_{topo}$ attributed to
the conserved current 
$J^\mu=(1/2\pi)\epsilon^{\mu\nu\lambda}\partial_\nu A_\lambda$. 
However,
there exists
stable, static and singular gauge field configurations 
which carry $q$ units of the $\rm U(1)_{\rm topo}$ charge.
These are the monopole operators that spontaneously
breaks the topological symmetry to create $4\pi q$
magnetic flux locally while satisfying the Dirac quantization 
constraint $2q\in\mathbb{Z}$ \cite{polyakov77}.
Although these are local operators, they
can not be constructed as polynomials of the 
fundamental fields of the theory, which makes it difficult to calculate
their scaling dimension using direct methods of Feynman diagrams.
However, the 
$e^2\rightarrow\infty,\ N\rightarrow\infty$ fixed point of ${\rm\ cQED}_3$ is 
conformal, and for conformal field theories (CFT) the scaling 
dimension of local operators can be determined using
state-operator correspondence of the radial quantization 
picture. 

For a $D$-dimensional quantum field theory, the usual quantization 
has the Hilbert space of states defined on a $(D-1)$-dimensional 
subspace with the remaining 
direction 
involving the time evolution generated by the Hamiltonian. In 
radial quantization the states are defined on concentric 
$(D-1)$-dimensional 
spheres of varying radii with the radial evolution generated by the dilatation 
operator $D=-ix^\mu\partial_\mu$.  It further follows that 
a local operator $O$ of a CFT 
inserted at the origin of
flat $\mathbb{R}^3$ space-time has a one-to-one correspondence to
normalizable states of the CFT on $S^2\times\mathbb{R}$. For a CFT 
with a trivial vacuum $|0\rangle$, it can be shown that a state 
$|\Delta_{\mathcal O}\rangle $, created by a conformal primary operator 
$\mathcal O(0)|0\rangle=|\Delta_{\mathcal O}\rangle $ at the origin  
is an eigenstate of the dilatation generator \cite{rychkov16},
\beq
D|\Delta_{\mathcal O}\rangle
=i\Delta_{\mathcal O}|\Delta_{\mathcal O}\rangle.
\eeq
Following a cylindrical transformation $\tau=\ln r$, it is easy to see that 
the dilatation generator plays the role of the Hamiltonian for such 
radial states. Therefore, the scaling dimension
$\Delta_{O}$ of the operator on $\mathbb{R}^3$ is equal to 
the energy of the corresponding
state on $S^2\times\mathbb{R}$. In this scenario
the energy eigenvalue of the state corresponding to 
the monopole operator $\mathcal{M}^{(q)}$ of charge $q$ at the origin
amounts diagonalizing
the $\rm\ cQED_{2+1}$ action on $S^2\times\mathbb{R}$ in the presence 
of $4\pi q$ unit of magnetic flux \cite{borokhov02}. However, the 
$\rm\ cQED_{2+1}$ action can only be diagonalized in the large-$2N$ limit 
with all fluctuations suppressed. For perturbations around the large-$2N$
limit, the ground state (free) energy $F^{(q)}=-\ln Z^{(q)}_{S^2\times\mathbb R}$ of 
the flux inserted action has to be computed order by order
such that the scaling dimension is obtained as \cite{pufu14},
\begin{equation}
\begin{aligned}
\Delta_{\mathcal{M}^{(q)}}&=F^{(q)}-F^{(0)}, \text{ with }\\
F^{(q)}&=-\ln Z^{(q)}_{S^2\times\mathbb{R}}
=-\lim_{\beta\rightarrow\infty}
\frac{1}{\beta}\ln Z^{(q)}_{S^2\times S^1_\beta}
.
\end{aligned}\label{eq:partition_sum}
\end{equation}
In the last line we have interpreted the groun state energy as the zero temperature
limit ($\beta\rightarrow\infty$) of the free energy where all the spatial and 
temporal directions are compact
\cite{dupuis19}. The above expression for the
scaling dimension subtracts a potentially divergent background 
free energy in the absence of any monopoles which does not affect physical 
quantities.

We have to consider the ${\rm\ cQED}_{2+1}$ action in the curved 
$S^2\times\mathbb{R}$ space-time. From the Euclidean signature the 
vierbein $\tensor{e}{_\mu^a}$ can be introduced to get a curved
space metric $g_{\mu\nu}=\tensor{e}{_\mu^a}\tensor{e}{_\nu^a}$. 
Eliminating any spin-connection by performing appropriate unitary 
rotation, the ${\rm\ cQED}_{2+1}$ action in the curved space-time
can be written as   
\begin{equation}
    \begin{aligned}
        S_{\rm\ cQED}
        =\int d^3r \sqrt{g} \
        \bar{\psi} \
        \tensor{e}{^\mu_a}
            \gamma^a
        \left[
            \partial_\mu
            +iA_\mu
            \right]
        \psi,
    \end{aligned}
\end{equation}
where $\gamma^a$ are the three spinor matrices defined on the
flat space time and $\sqrt{g}$ is 
the short-hand for the square-root of 
the metric determinant $\sqrt{\det g_{\mu\nu}}$. It is introduced to define the 
coordinate invariant volume measure $d^3r\sqrt{g}$. 
Insertion of a monopole of charge $q$ amounts to embedding 
a $q$ unit of magnetic flux at the origin by introducing a 
singular gauge field configuration. The static gauge field 
contribution due to the monopole at the center is 
$
\vec{\mathcal{A}}_q=\frac{q}{2}\frac{1-\cos\theta}
{r\sin\theta}\hat{e}_\phi
$. Mapping to the cylindrical space-time 
$\mathbb{S}^2\times\mathbb{R}$ with the metric 
$ds^2=g_{\mu\nu}dr^\mu dr^\nu=d\tau^2+(d\theta^2+{\sin\theta}^2d\phi^2)$
from the usual spherical co-ordinates in 
$\mathbb{R}^3$ with the metric 
$ds^2=dr^2+r^2(d\theta^2+{\sin\theta}^2d\phi^2)$
is obtained by putting $r=e^\tau$ and performing a Weyl 
rescaling,
\begin{equation}
    \begin{aligned}
        g_{\mu\nu}\rightarrow e^{-2\tau}g_{\mu\nu},& \ \  
        \psi,\bar{\psi}\rightarrow 
        e^{-\tau}\psi,e^{-\tau}\bar{\psi},
        \\
        \tensor{e}{^\mu_a}\rightarrow e^{-\tau} 
        \tensor{e}{^\mu_a},& \ \  
        A_\mu\rightarrow A_\mu.&
    \end{aligned}
    \label{eq:weyl_rescaling}
\end{equation}
The transformed Dirac operator in the presence of a magnetic monopole
of charge $q$ in the cylindrical space-time is given by 
\cite{borokhov02} 
\begin{equation}
    \begin{aligned}
        \slashed{D}
        =\gamma_r
        \left[\frac{\partial}{\partial\tau}
        -\left(J^2-L^2+\frac{1}{4}\right)
        +q\gamma_r
        \right],
    \end{aligned}\label{eq:monopole_dirac}
\end{equation}
where $\gamma_r=\hat{r}\cdot\vec{\gamma}$. Here, $\vec{J}$ and
$\vec{L}$ are the generalized total and orbital
angular momenta, respectively, in the presence of the monopole magnetic flux.
At this level dynamical contribution towards the gauge fields
are ignored. This is strictly valid in the large-$N$ limit 
and their sub-leading effect can be incorporated back within a controlled 
$1/(2N)$ expansion \cite{pufu14}.

Following earlier work by Ref.~\onlinecite{wu76},
it was shown by Ref.~\onlinecite{borokhov02} that the 
Dirac operator in the presence of a monopole generated
background gauge field can be diagonalized by a
special monopole harmonics basis. In the presence of a monopole of 
charge $q$, the monopole harmonics are 
defined as $L^2Y_{q,lm}=l(l+1)Y_{q,lm}$, 
$L_zY_{q,lm}=mY_{q,lm}$ with $l=|q|,|q|+1,|q|+2,\dots$
and $m=-l,\dots,l$. The Dirac equation is not 
diagonal in the monopole harmonics basis. Instead a basis
involving two separate modes of the total angular momentum 
$j=l\pm\frac{1}{2}$ needs to be considered
\begin{equation}
    \begin{aligned}
        T_{q,lm}(\theta,\varphi)
        &=\begin{pmatrix}
            \sqrt{\frac{l+m+1}{2l+1}}Y_{q,lm}
            (\theta,\varphi)\\
            \sqrt{\frac{l-m}{2l+1}}Y_{q,l(m+1)}
            (\theta,\varphi)
        \end{pmatrix} : j = l + \frac{1}{2}, \\
        S_{q,lm}(\theta,\varphi)
        &=\begin{pmatrix}
            -\sqrt{\frac{l-m}{2l+1}}Y_{q,lm}
            (\theta,\varphi)\\
            \sqrt{\frac{l+m+1}{2l+1}}Y_{q,l(m+1)}
            (\theta,\varphi)
        \end{pmatrix} : j = l - \frac{1}{2},
    \end{aligned}
    \label{eq:harmonic_basis}
\end{equation}
which brings the monopole Dirac equation to an almost
diagonal form. Following the notation of Ref.~\onlinecite{pufu14} 
we can write down the $2\times 2$ eigenvalue equation 
of the Dirac operator in the basis 
$\left(T_{q,(l-1)m},S_{q,lm}\right)^T$,
\begin{equation}
    \begin{aligned}
        \slashed{D}
        \begin{pmatrix}
            T_{q,(l-1)m}e^{-i\omega\tau}\\
            S_{q,lm}e^{-i\omega\tau}
        \end{pmatrix}
        =d_{q,l}(\omega)
        \begin{pmatrix}
            T_{q,(l-1)m}e^{-i\omega\tau}\\
            S_{q,lm}e^{-i\omega\tau}
        \end{pmatrix},
    \end{aligned}\label{eq:eigen}
\end{equation}
where $d_{q,l}(\omega)=A_{q,l}(-i\omega+B_{ql})$ is the 
eigenvalue matrix given by
\begin{equation}
    \begin{aligned}
        A_{q,l} &=
        \begin{pmatrix}
            -\frac{q}{l}&&-\sqrt{1-\frac{q^2}{l^2}}\\
            -\sqrt{1-\frac{q^2}{l^2}}&&\frac{q}{l}
        \end{pmatrix},\\
        B_{q,l} &=
        \begin{pmatrix}
            l\left(1-\frac{q^2}{l^2}\right)
            &&-q\sqrt{1-\frac{q^2}{l^2}}\\
            -q\sqrt{1-\frac{q^2}{l^2}}
            &&-l\left(1-\frac{q^2}{l^2}\right),
        \end{pmatrix}
    \end{aligned}
\end{equation}
The monopole
harmonics are defined for $l\geq |q|$, and for the case involving
$l=q$ the matrices $A_{q,l}$ and $B_{q,l}$ are one dimensional
with the only term given by their bottom-right entry.
In this semi-diagonal basis, the zeroth-order ground-state energy is 
easily obtained by integrating out the fermions from the path 
integral, and we get
\begin{equation}
    \begin{aligned}
        F^{(q)}_0
        &=-\frac{1}{\beta}\Tr_{S^2\times S^1_\beta}\ln\left[\slashed{D}
        \right]\\
        &=-(2N)
        \int\frac{d\omega}{2\pi}
        \sum_{l=q}^\infty
        \sum_{m=-l}^{l-1}
        \ln\det
        \left[d_{q,l}(\omega)\right]\\
        &=-(2N)
        \int\frac{d\omega}{2\pi}
        \sum_{l=q}^\infty
        2l\ln\left(\omega^2+l^2-q^2\right)\\
        &=(2N)\lambda_0^{(q)},
    \end{aligned}\label{eq:clean_free}
\end{equation} 
where $\lambda^{(q)}_0$ is a regulated summation which can be
computed and has been tabulated for various $q$'s in 
Ref.~\onlinecite{pufu14}. After appropriate
regularization the 
expression obtained in Ref.~\onlinecite{pufu14} yields 
$\lambda^{(0)}_0=0$, so that the regulated free energy for 
$q>0$ is equal to the finite free (Casimir) energy difference 
of adding a $q$ charge monopole in $S^2\times\mathbb{R}$.
The dynamical gauge fields which have been ignored
so far can be introduced as loop corrections to the
above free energy. Roughly, this involves expanding
the full trace-logarithm 
$F^{(q)}=-\Tr\ln\left[\slashed{D}+i\slashed{A}\right]$ 
in the gauge field strength and integrating them out. Its
contribution is suppressed by a factor of $1/(2N)$, and
the scaling dimension of the monopole operators 
beyond the large $2N$ limit is, therefore, given by,
\begin{equation}
    \begin{aligned}
        \Delta_{\mathcal{M}^{(q)}}
        =(2N)\lambda_0^{(q)}+\lambda_1^{(q)}
        +O\left(1/(2N)\right)
    \end{aligned}
\end{equation}
The subleading correction $\lambda^{(q)}_1$ for the first
few charges $q$ are provided in Ref.~\onlinecite{dyer13}.
In this same spirit, we now consider random perturbations to
the ${\rm\ cQED}_{2+1}$ action and calculate the perturbative
correction to the monopole scaling dimension obtained in the
clean limit.

\section{Monopole scaling dimension of dirty ${\rm cQED}_{2+1}$}
\label{dirty_sdim}

We wish to include the effect of RC perturbation on the
monopole free energy obtained above within the large-$N$ 
perturbation theory. The application of state-operator correspondence 
requires two ingredients: (1) radial quantization where the Hilbert 
space is defined on concentric spheres and (2) conformal symmetry 
which ensures that the states corresponding to primary operators with well 
defined scaling dimensions at the origin are eigenstates of the dilatation 
generators. Furthermore, we need to map the radially quantized theory 
on a cylinder which requires Weyl invariance \eqref{eq:weyl_rescaling}
of the action. In principle there is no problem with using the radial 
quantization picture with Hamiltonians involving quenched disorder.   
Conformal symmetry is definitely absent for a particular realization of disorder.
However, we will consider the state-operator correspondence
only after performing the disorder averaging which restores the
homogeneity of space-time. The disorder averaged theory may lack conformal 
symmetry in the infrared limit, and this will be signalled by the 
unitarity violation of the primary operators of the theory \cite{mudry96}. 
Quenched randomness is static in time and if we seek to establish a 
connection with the radial quantization picture, such random couplings must
be parametrized by the co-ordinates on the two-sphere.  

In the standard treatments, quenched random couplings are parametrized
on the planar spatial $\mathbb{R}^2$ submanifold of the $(2+1)$-dimensional 
space-time manifold.
To obtain the correspondence between the disorder strengths of random 
couplings defined on a space-like plane and a space-like sphere we 
consider a one-point compactification of the two-dimensional plane,
\begin{equation}
    \begin{aligned}
        (x,y) &= (\tan\frac{\theta}{2}\cos\phi,\tan\frac{\theta}{2}\sin\phi),
    \end{aligned}
\end{equation}
which transforms the planar spatial metric $ds^2_\parallel=dx^2+dy^2$ 
into $ds^2_\parallel=\frac{1}{4}{\sec\frac{\theta}{2}}^4d\theta^2
+{\tan\frac{\theta}{2}}^2d\phi^2$. Naturally, the induced metric on 
the sphere $ds^2_\parallel$ differs from the usual spherical metric 
$ds^2_{S^2}=d\theta^2+\sin^2\theta d\phi^2$. In the compactified
space the Gaussian weight for the random couplings in 
\eqref{eq:dis_pert}, therefore, becomes 
\begin{equation}
    \begin{aligned}
        P[V]=\exp\left(-\frac{1}{2\rho_\alpha}\int d^2r_\parallel
        \sqrt{g_\parallel}
        V_{\alpha j}^2
        \right).
    \end{aligned}
    \label{eq:mod_dist}
\end{equation}
Here, following our notation, $d^2r_{\parallel}\sqrt{g_{\parallel}}$ is the 
invariant volume measure of the compactified sphere, but 
$\rho_\alpha$'s are the same disorder strengths as defined on the flat space. 
Through this transformation,
we have mapped the content of the physical information about the quenched 
disorder, 
i.e., the autocorrelation of the random couplings into a radially quantized theory. 
Once we are able to calculate the scaling dimension of the monopole operator of this
theory we can transform back to usual $\mathbb R^3$ space-time and retrieve 
the standard on-site autocorrelation of random couplings defined on the flat space. 

In the $S^2\times\mathbb{R}$ space, the disordered perturbation to the 
clean ${\rm\ cQED}_{2+1}$ action is given by
\begin{equation}
    \begin{aligned}
        S_{\rm dis}
        =\int d^3r\sqrt{g}
        \bar{\psi}
        \left(i V_{\alpha j}\sigma^\alpha 
        e^j_a\gamma^a
        \right)
        \psi.
    \end{aligned}
\end{equation}
The Weyl rescaling \eqref{eq:weyl_rescaling} leaves the RC fields 
$V_{\alpha j}$ unchanged. This distinguishes them from random
mass perturbations (which do not stay invariant under the Weyl
rescaling) and allows us to move forward with the 
radial quantization technique. For random mass-type disorder 
the Weyl rescaling leads to a scaling of the random 
coupling field $M\rightarrow e^{-\tau}M$ 
and introduces infrared divergences to the theory. 
With the chosen disorder distribution \eqref{eq:mod_dist}, 
the on-site correlation between the random couplings is given by,
$
\overline{V_{\alpha j}(\theta,\phi)V_{\beta k}(0,0)}
        =\rho_\alpha\delta_{\alpha\beta}
        \delta_{jk}\delta(\theta)\delta(\phi)/
        \sqrt{g_\parallel}
$. In the integrated form it yields
\begin{equation}
    \begin{aligned}
\int d^3r\sqrt{g}
f(\tau)
\overline{V_{\alpha j}(\theta,\phi)V_{\beta k}(0,0)}
        =4\int d\tau f(\tau)\rho_\alpha\delta_{\alpha\beta}\delta_{jk},
    \end{aligned}
\end{equation}
here formally the $\tau$ is integrated from $0$ to $\beta$ in the temporal 
direction of the $S^2\times\mathbb{R}$ cylindrical space.
This additional factor of $4$ 
establishes a correspondence between the strengths of the 
random couplings 
on the sphere and the plane. After we extract the scaling dimension 
of a single monopole operator inserted at the center in the disordered 
background we will consider the situation of a dilute concentration 
of monopole gas \eqref{eq:monopole_gas} in the presence of quenched 
disorder in the usual space-time. In the limit of vanishing dilution 
of the monopole operators, we will 
use the known expression \cite{thomson17,zhao17} for the
RG flow of the random couplings in the non-compact disordered 
$\rm\ QED_{2+1}$ and disorder modified RG flow of the monopole fugacity 
to fully characterize the disorder driven instabilities of the compact 
theory. The Coulomb gas of the monopoles does modify the renormalization-group 
flow of the minimally coupled 
gauge charge in the matter sector 
\cite{hermele04,nogueira08}, but to the 
leading order considered here, 
the gauge charge renormalization (and, therefore, the renormalization of 
the monopole fugacity) does not contribute to the renormalization of 
random couplings \cite{thomson17}.
It is to be noted that the gauge charge flows to strong coupling 
in this context \cite{nogueira08,thomson17,zhao17} and can be left out of the RG 
flow equations.
Following disorder averaging the 
scaling dimension of the monopole operator can be simply extracted 
from the difference of the free energies, 
$
\Delta_{\mathcal{M}^{(q)}}
        =\overline{F^{(q)}(V_{\alpha j})}
        -\overline{F^{(0)}(V_{\alpha j})}
$. 

As the RCs couple quadratically to fermionic
fields, we can formally integrate out the fermions from the generic 
disordered ${\rm\ cQED}_{2+1}$ action, and similar to what is
performed 
with 
the dynamical gauge fields, perturbatively expand the resulting 
expression in random coupling strength and perform direct disorder 
averaging. It is convenient to exploit the homogeneity of 
the $S^2\times S^1_\beta$ space-time post disorder 
averaging and compute the functional trace in the space-time basis
as, $\Tr_{S^2\times S^1_\beta}\overline{A}
=\mathcal{V}(S^2)\mathcal{V}(S^1_\beta)
\
\tr
\langle r_0|
\overline{A}|
r_0 \rangle
$,
where $\mathcal{V}$ denotes the volume of the space,
$r_0$ is any given point in the space, and $\tr$ is a trace
within the Dirac spinor and SU(2N) flavor space. In the following we 
choose the north pole co-ordinates $r_0=(\tau=0,\theta=0,\phi=0)$
and obtain, 
\begin{widetext}
    \begin{equation}
        \begin{aligned}
            &F^{(q)}(\rho_\alpha)
                        =
                        -\overline{\Tr_{S^2\times S^1_{\beta}}\ln\left[
                            \slashed{D}
                            +i\slashed{A}
                        +iV_{\alpha j}\sigma^\alpha e^j_a\gamma^a\
                        \right]}
                        +O(1/(2N))
                        \\
                    =&
                    (2N)\lambda^{(q)}_0+\lambda^{(q)}_1
                        +\lim_{\beta\rightarrow\infty}
                        \frac{\mathcal{V}(S^2)\mathcal{V}(S^1_\beta)
                        }{2\beta}
                        \int d^3r\sqrt{g} \
                        \overline{\tr\left[
                            G^{(q)}(r_0,r)
                        i V_{\alpha j}(\theta,\phi) 
                        \sigma^\alpha e^j_a\gamma^a
                        G^{(q)}(r,r_0)
                        i V_{\beta k}(0,0)
                        \sigma^\beta e^k_b\gamma^b
                        \right]}
                    +O\left(1/(2N),V^4\right)\\
                    =&
                    (2N)\lambda^{(q)}_0+\lambda_1^{(q)}
                        +2\pi(4\rho_\alpha)
                        \int d\tau \
                        \tr\left[
                            G^{(q)}(\tau)
                        (i\sigma^\alpha e^j_a\gamma^a)
                        G^{(q)}(-\tau)
                        (i\sigma^\alpha e^j_b\gamma^b)
                        \right]
                    +O(1/(2N),\rho_\alpha^2),
        \end{aligned}\label{eq:free_energy}
    \end{equation}
\end{widetext}
where $G(\tau)$ is the monopole Green's function 
between coincident angles
$G(\tau)=\langle r_0|\slashed{D}^{-1}|r\rangle$ with $r=(\tau,0,0)$.
For a controlled perturbative double expansion we must have, 
$\rho_\alpha \sim 1/(2N)$. In that way, following the 
trace over the vertex matrices, the first order disorder 
contribution is an $O(1)$ perturbation to the zeroth-order 
free energy \eqref{eq:clean_free}. The spectral 
decomposition of the Green's function matrix is expressed in the 
$2\times 2$ monopole spherical harmonic basis \eqref{eq:harmonic_basis} as
\begin{widetext}
\begin{equation}
    \begin{aligned}
        G(\tau)
        =
        \left. \int\frac{d\omega}{2\pi}
        e^{-i\omega\tau}
        \sum_{l=q}^\infty\sum_{m=-l}^{l-1}
        \begin{pmatrix}
            T_{q,l-1,m}&
            S_{q,l,m}
        \end{pmatrix}
        d_{q,l}(\omega)^{-1}
        \begin{pmatrix}
            T^\dagger_{q,l-1,m}\\
            S^\dagger_{q,l,m}
        \end{pmatrix}
        \right|_{r_0}.
    \end{aligned}
\end{equation}
\end{widetext}
The full expression of the Green's function is given in 
Ref.~\onlinecite{pufu14} and involves complicated special functions.
The present scenario, however, is simpler and 
at the north pole using
the property  
$Y_{q,lm}=\delta_{q,-m}\sqrt{(2l+1)/(4\pi)}$ \cite{dyer13},
we have a much simpler expression for the coincident angle
Green's function,
\begin{equation}
    \begin{aligned}
        G(\tau)=-\frac{{\rm sgn}(\tau)}{2}
        \left[
            \frac{q}{4\pi}
            \left(I+\gamma^0\right)
            +\sum_{l=q+1}^\infty
            \frac{l}{2\pi}
            e^{-\sqrt{l^2-q^2}|\tau|}\gamma^0
            \right],
    \end{aligned}\label{eq:rspace_greens}
\end{equation}
where $I$ is a $2\times 2$ identity matrix.
In the contribution to the free energy, the Green's function convolution 
includes a summation over the Dirac matrices. Evaluated at the north pole
the summation product of the Dirac matrices yields
$(i\gamma^1)\gamma^0(i\gamma^1)+(i\gamma^2)\gamma^0(i\gamma^2)
=2\gamma^0$ and
$(i\gamma^1)I(i\gamma^1)+(i\gamma^2)I(i\gamma^2)
=-2I$.
Additionally, the summation and trace 
over the SU(2N) matrices contributes a numerical pre-factor
$=2\times\tr\left[\sigma^\alpha
\sigma^\alpha \times 1_{N\times N}\right] = 4N$. In the following,
we will also abbreviate $\sum_\alpha\rho_\alpha =\rho$.

The singular piece of the free energy is determined from the 
short-distance ultra-violet (UV) behavior of the Green's function 
[Eq.~\eqref{eq:rspace_greens}]. In its written form a small 
$|\tau|$ expansion of Eq.~\eqref{eq:rspace_greens} is not very useful.
Instead,
we expand the expression in small $q$ and perform the resulting
convergent sums over $l$ 
directly (see Ref.~\onlinecite{dyer15}) and then
perform an asymptotic expansion to find,
\begin{widetext}
\begin{equation}
    \begin{aligned}
        G(\tau)
        =
        -\frac{{\rm sgn}(\tau)}{2}\left[
        \frac{q}{4\pi}
        1
        +
        \left(\frac{1}{2\pi\tau^2}
        -\frac{1}{24\pi}
        +\frac{q(8+q(12+q(-8+\pi^2q)))|\tau|}{96\pi}
        +\frac{(1-20q^2)\tau^2}{480\pi}
        +O(q^5,\tau^3)
        \right)\gamma^0\right].
    \end{aligned}\label{eq:short_greens}
\end{equation}
\end{widetext}
It is simpler to track the UV contribution of the free energy by going to
the frequency space.
With the Fourier transformations,
${\rm F.T}[{\rm sgn}(\tau)]=-\frac{2}{i\omega}$, 
${\rm F.T}[\frac{1}{\tau^2}]=-\pi|\omega|$, 
${\rm F.T}[|\tau|]=-\frac{2}{\omega^2}$ and so forth, it follows
that in the given order in the perturbation theory the only terms which
depend on an UV frequency cutoff $\Lambda$ in the free energy 
[Eq.~\eqref{eq:free_energy}] are independent of $q$. The $q$-independent 
singular pieces are unimportant as they drop out from the contribution 
to the scaling dimension, $\Delta_{\mathcal{M}^{(q)}}
=\overline{F^{(q)}(\rho)}-\overline{F^{(0)}(\rho)}$. As the non-analytic 
portion of the Green's function is independent of $q$, the inference 
about the UV properties obtained from the current asymptotic
expansion holds irrespective of at which order of $q$ we truncate
the above series.

Having established that the quenched RC perturbation
does not introduce any physical UV singularities we now set out to 
extract the finite part of the free energy. After performing the
time integral and taking the trace over the Dirac and SU(2N) matrices 
we obtain
\begin{widetext}
    \begin{equation}
        \begin{aligned}
            \overline{F^{(q)}(\rho)}
            &=(2N)\lambda^{(q)}_0+\lambda_1^{(q)}
            -\frac{2\rho (2N)}{\pi}
            \left[
                q\sum_{l=q+1}^\infty\frac{l}{\sqrt{l^2-q^2}}
            +
            \sum_{l,l'=q+1}^\infty\frac{ll'}
            {\sqrt{l^2-q^2}+\sqrt{{l'}^2-q^2}}
            \right]
            +O(1/(2N),\rho^2_\alpha).
        \end{aligned}\label{free_here}
    \end{equation}
\end{widetext}
It is to be noted that the constant $\tau$-independent
piece of the Green's function 
[Eq.~\eqref{eq:rspace_greens}] does not contribute a constant 
contribution
to the integrand, and, therefore, the convolution does not lead
to any infrared singularities.
This is particular to the case of the random non-Abelian vector potential
perturbation that we have considered. Indeed,
e.g., for a random scalar potential which has the Dirac matrix
 $\tensor{e}{^r_a}\gamma^a$ on $S^2\times\mathbb{R}$ ($\gamma^0$ in the
 flat space-time) in its vertex, 
the contribution from the constant piece does not drop out.

For the present case we have a disordered correction for the 
free energy [Eq.~\eqref{free_here}] with formally divergent
summations over the quantum numbers of the 
monopole harmonic basis \eqref{eq:harmonic_basis}. The zeroth-order 
contribution to the free energy \eqref{eq:clean_free} has 
been formally obtained by using the $\zeta$-function regularization 
technique \cite{pufu14}. We employ the same method for the 
disorder correction term. With $\zeta$-function regularization, the 
finite part of an apparently divergent infinite series 
summation can be obtained using analytic 
continuation of the Hurwitz $\zeta$ function \cite{NIST:DLMF},
\begin{equation}
    \begin{aligned}
        \sum_{l=0}^\infty
        (l+z)^{s}=\zeta(-s,z)=-\frac{B_{s+1}(z)}{s+1}
        \ \
        \forall \ \ s\neq 1.
    \end{aligned}
\end{equation}
For the first sum, 
$I_1(q)=q\sum_{l=q+1}^\infty l/\sqrt{l^2-q^2}$ we note that the
summand remains $\propto 1$ for large $l$. The regularized summation
can be obtained by
considering a different quantity, 
$l/\left(l^2-q^2\right)^{s/2}$ for an $s$ where the summation
is perfectly convergent, and then the result can be 
analytically continued to $s=1$. This is a well known technique
to regularize the free energies of radially quantized
CFTs \cite{pufu13,pufu14,dupuis19,dyer15,dyer13} and one of the central 
computational elements for treating spherically symmetric problems in
similar contexts.
For this purpose we need to add and subtract the 
asymptotic form of the summand and obtain
\begin{equation}
    \begin{aligned}
        I_1(q)/q&=
        \lim_{s\rightarrow 1}\left[\sum_{l=q+1}^\infty
        \left(\frac{l}{\left(l^2-q^2\right)^{s/2}}
        -l^{1-s}
        \right)
        +\sum_{l=q+1}^\infty l^{1-s}\right],\\
        &=R_1(q)+\zeta(0,q+1),
    \end{aligned}
\end{equation}
where in the second step we can take the limit $s=1$ by using
the Hurwitz $\zeta$ function identity for the formally
divergent term. Here $R_1(q)$ is a perfectly convergent summation 
which can be evaluated up to arbitrary numerical accuracy. The same 
technique can be extended to obtain the finite contribution from the
second sum, $I_2(q)=\sum_{l,l'=q+1}^\infty
\frac{ll'}{\sqrt{l^2-q^3}+\sqrt{{l'}^2-q^2}}$, but owing to the
presence of a double summation, the resulting expression is
cumbersome. The complete expression for the regularized double summation
contribution with an unimportant $q$-independent piece subtracted
$\tilde{I}_2(q)=I_2(q)-I_2(0)$ has been provided in the Appendix. 

\begin{figure}
  \includegraphics[width=0.8\columnwidth]{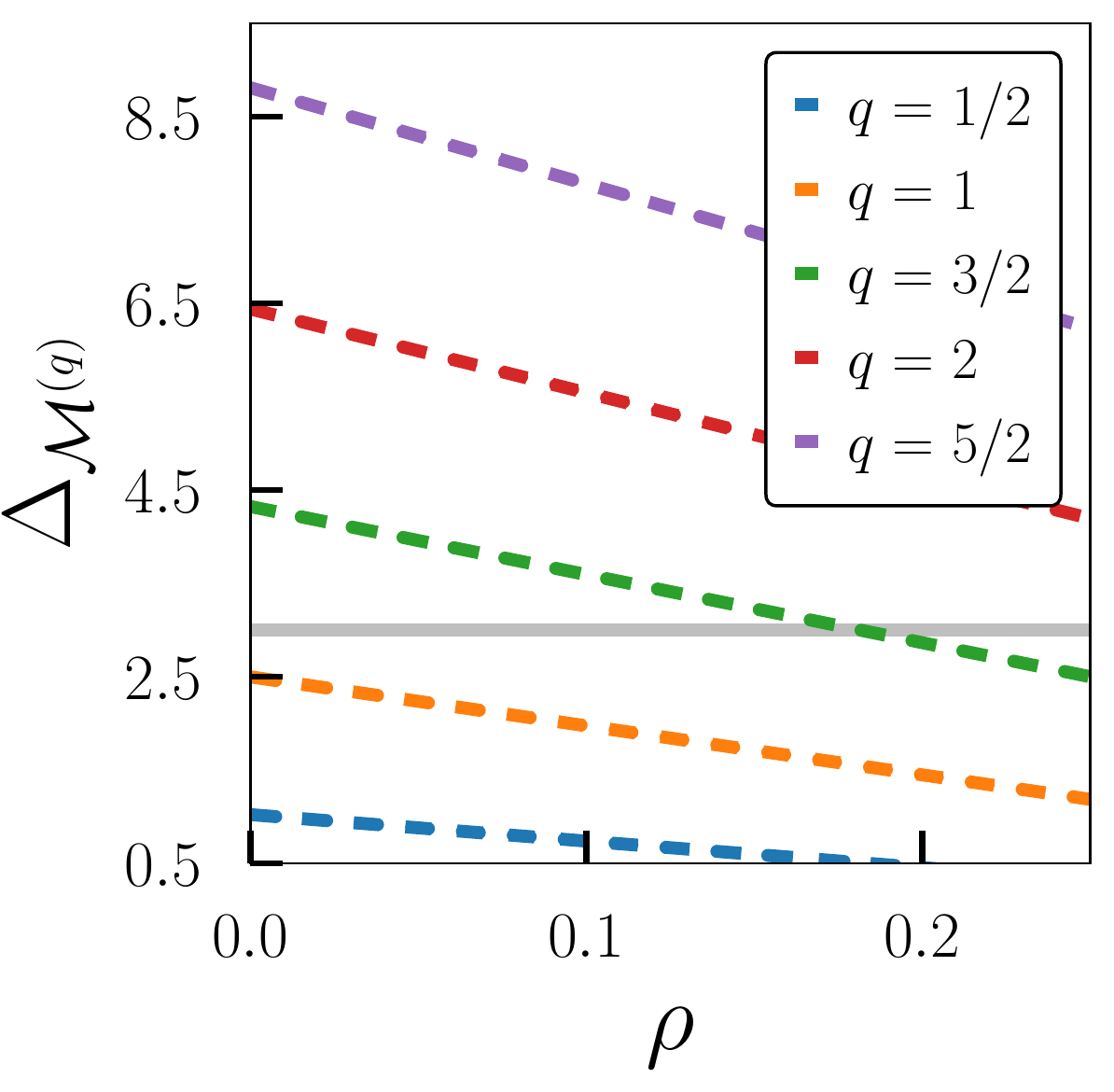}
 \caption{
     The bare scaling dimension of monopole operators 
     [Eq.~\eqref{eq:scaling_dimension}]
     as a function of bare random current
     strength $\rho$ with $(2N)=4$
     flavors of fermions. The gray line denotes the threshold minimum
     $\Delta_{\mathcal{M}^{(q)}} = 3$, below which the monopoles
     proliferate. In the triangular lattice 
     staggered $\pi$-flux spin-liquid state which only allows
     monopole operators of charges 
     $q\geq 3/2$, disorder reduces the scaling dimension 
     of monopoles of all charges.
 }
 \label{fig:scaling_dimension}
\end{figure}

From the disorder averaged free energy the bare scaling dimension of the
monopole operator of charge $q$ is therefore given by
\begin{equation}
    \begin{aligned}
        \Delta_{\mathcal{M}^{(q)}}
        =&(2N)\lambda^{(q)}_0
        +\lambda_1^{(q)}
        -\frac{2\rho (2N)}{\pi}
        \left[I_1(q)+\tilde{I}_2(q)\right]\\
        &+O(1/(2N),\rho_\alpha^2),
    \end{aligned}\label{eq:scaling_dimension}
\end{equation}
where the disorder contribution
 $\rho=\sum_\alpha\rho_\alpha$ is to be summed over the
SU(2) indices $\alpha$, depending on the residual symmetries of the
RC disorder. The unitarity bound 
dictates that the scaling dimension of a conformal scalar operator has 
to be $\geq 0.5$ in $D=2+1$. It clearly follows that at the 
critical disorder strength $\rho^*=
\pi[(2N)\lambda^{(q)}_0+\lambda^{(q)}_0-0.5]/\{2(2N)[I_1(q)+\tilde{I}_2(q)]\}$, 
the 
bound is saturated for the monopole operator of charge $q$. This 
signals the breakdown of the conformal symmetry of the disordered 
infrared fixed point.

However, more importantly,  it is
the RG relevance of the monopole fugacity operator $y^{(q)}$ , 
which has the scaling dimension
$3-\Delta_{\mathcal{M}^{(q)}}$ \eqref{eq:fugacity_rg} 
in three space-time dimensions
that dictates the suppression or
proliferation of the monopoles \cite{hermele04,tong18}. 
From Eq.~\eqref{eq:scaling_dimension} we see that the presence of 
disorder under-screens the monopole 
operator such that the monopole fugacity is made more relevant. This is an
important finding of our paper.
In the triangular lattice 
staggered $\pi$-flux Dirac spin-liquid phase the allowed monopole charges are 
$q\geq 3/2$. Considering generic RC perturbation 
with disorder strength $\rho=\sum_{\alpha}\rho_\alpha$,
it clearly emerges from 
Fig.~\ref{fig:scaling_dimension} that
with increasing disorder strength, higher charged 
monopole fugacities become relevant but the instability is 
still instigated by the 
proliferation of the monopoles of the lowest allowed charge $q=3/2$. In the 
large-$2N$ limit the $q=1$ monopole in the kagome lattice is allowed 
as a composite operator along with the spinons \cite{song19}. 
Although this is already a 
relevant operator in the clean limit, from Fig.~\ref{fig:scaling_dimension} it clearly 
emerges that finite RC perturbation makes such a composite operator even more 
relevant. The proliferation of the monopoles leads to the confinement of 
fractionalized spinons of the destruction of the spin-liquid phase.
For generic couplings to random SU($2N$) currents, the disorder
strengths $\rho_\alpha$ also renormalize \cite{thomson17,zhao17,goswami17}. In
that case the fate of the conformal fixed point is governed by
the combined renormalization of all the couplings of the problem.

\section{Disordered RG flow with monopoles}
\label{rg_flow}

To study the fate of the deconfined fixed point of the 
${\rm\ cQED}_{2+1}$ action, the renormalization of all the associated 
couplings and the related instabilities need to be considered
together \cite{hermele04}. In our case, this boils down to
the renormalization group flow of the monopole fugacity $y^{(q)}$ 
(which has bare scaling dimension 
$3-\Delta_{\mathcal{M}^{(q)}}$) and the random disorder couplings. The 
infrared flow of the gauge electric charge $e^2\rightarrow\infty$ is to 
the leading order unaffected by the disorder couplings 
\cite{nogueira08,thomson17} and can be ignored for the present discussion.

\begin{figure}
\includegraphics[width=0.8\columnwidth]{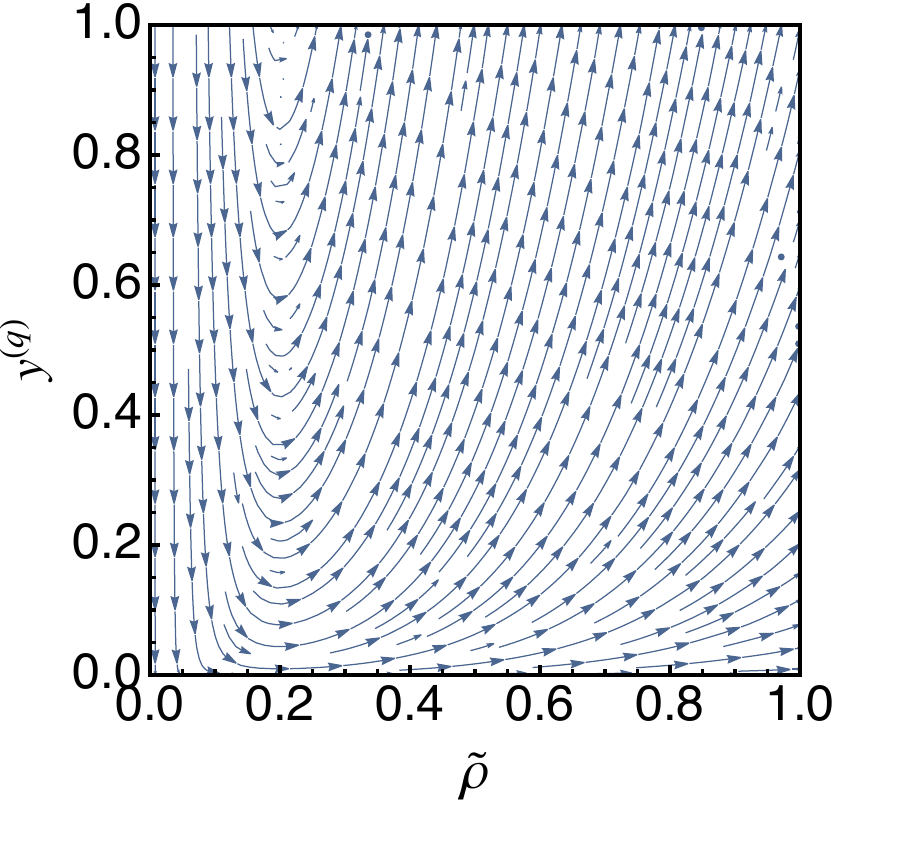}
 \caption{
     The combined RG flow of SU($N$) symmetric random current 
     coupling 
     $\rho_x=\rho_y=\rho_z=\tilde\rho$ and monopole fugacity 
     $y^{(q)}$ of the $q=3/2$ monopole operator which is 
     allowed for the triangular lattice staggered $\pi$-flux 
     U(1) Dirac spin-liquid 
     state.
 }
\label{fig:flow}
\end{figure}

The RG flow equation of RC couplings 
$\rho_\alpha$ was obtained in Ref.~\onlinecite{thomson17}
and in the absence any other form of disorder we can
adapt their expression to write,
$\frac{d\rho_\alpha}{dl}=
(4/\pi)|\epsilon_{\alpha\beta\gamma}|
\rho_\beta\rho_\gamma
$ where $\epsilon_{\alpha\beta\gamma}$ is the usual Levi-Civita
tensor. Compared to Ref.~\onlinecite{thomson17} we have the opposite 
sign convention of the RG flow, and our random current strengths 
are defined as twice of theirs (see Eqs.~(24), (34), and (44) 
from Ref.~\onlinecite{thomson17} for comparison).
This equation describes a flow of the genetic disorder 
strength to its strong-coupling fixed point, and, therefore, such 
random couplings introduce instability to the 
conformal fixed point of ${\rm\ cQED}_{2+1}$. Including
monopoles in the picture we have, to the leading order,
a combined RG flow equation,
\begin{equation}
    \begin{aligned}
        \frac{dy^{(q)}}{dl}&=\left(3-\sum_\alpha
        \Delta_{\mathcal{M}^{(q)}}(\rho_\alpha)
        \right)y^{(q)},\\
        \frac{d\rho_\alpha}{dl}
        &=\frac{4|\epsilon_{\alpha\beta\gamma}|}{\pi}
        \rho_\beta
        \rho_\gamma,
    \end{aligned}
\end{equation}
where the contributions to the scaling dimension of the monopole
operator from all the disorder couplings have been added 
together. The presence of all three $\rho_\alpha$ couplings would
 indicate that the
emergent SU($2N$) flavor symmetry has been broken down to a reduced 
SU($N$). 
From the coupled flow equation it is clear that the 
SU($N$) symmetric disorder moves to a strong coupling 
fixed point. Consequently, due to its linear regressive dependence on the 
disorder strength, the monopole fugacity 
also flows to strong coupling (see Fig.~\ref{fig:flow}). 
This observation leads to the clear indication that 
the Dirac spin-liquid phase is destroyed by SU($N$) symmetric 
RC disorder as magnetic monopoles
proliferate, and confinement ensues. From Fig. 
\ref{fig:scaling_dimension} we understand that the fugacity of
the monopoles with the lowest microscopically allowed charge 
turns relevant first as the disorder strength flows to its 
strong-coupling limit.

\begin{figure}
\includegraphics[width=0.8\columnwidth]{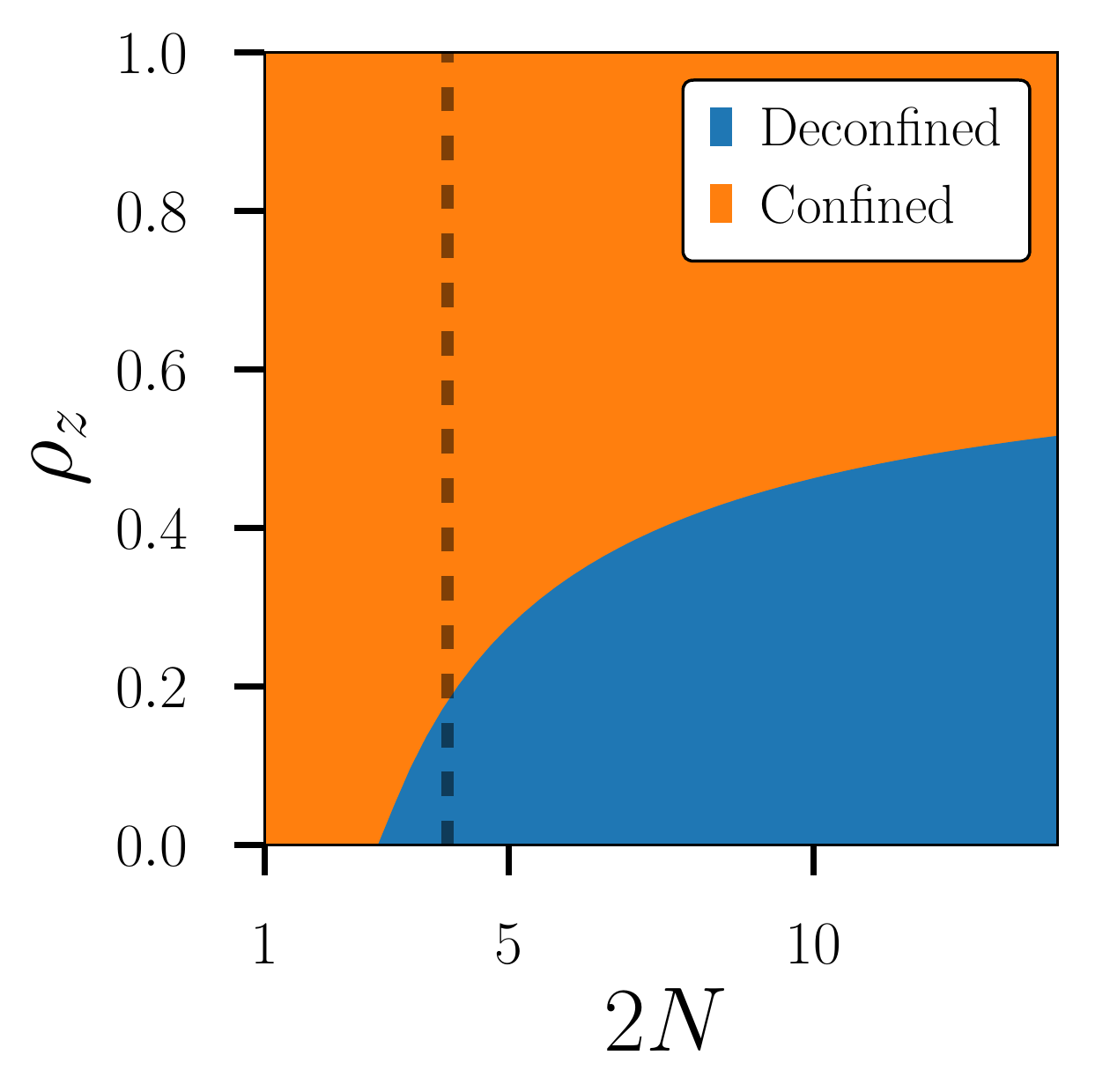}
 \caption{
     The phase diagram of the U(1) Dirac spin-liquid
     with $2N$ flavors of fermions coupled to RG marginal 
     $\rm U(1)\times SU(N)$ symmetric
     random current perturbations when only the monopole operators
     with charge $q\geq 3/2$ are symmetry allowed. In the triangular lattice
     the microscopic lattice
     symmetries disallow monopoles of charge $q<3/2$ for the 
     staggered $\pi$-flux ansatz.
     The gray line indicates the case of the triangular lattice
     staggered $\pi$-flux Dirac spin-liquid state which has $(2N)=4$.
    }
\label{fig:pd}
\end{figure}

Ref.~\onlinecite{thomson17} further showed that if the random
perturbations to the action obey more symmetries, the effect
of disorder may be less drastic. Particularly, for a random 
coupling $\rho_z$ to only one of the three components of SU(2) the
subgroup of SU($2N$) vector currents ($\rho_x=\rho_y=0$), 
the RC is $\rm U(1)\times SU(N)$ symmetric, and in 
this case, following the same RG equation from above, it turns out 
that the disorder coupling $\rho_z$ is marginal under RG. 
However, as demonstrated in Fig.~\ref{fig:pd}, for the case of 
the triangular lattice where monopole operators of charge $q<3/2$ 
are prohibited, a confinement transition ensues at a finite 
disorder strength and the spin-liquid 
phase is destabilized at a finite critical value of 
$\rho_z^c \sim 0.175$. 
As per our definition of the RC
perturbation [Eq.~\eqref{eq:dis_pert}], this critical disorder 
strength is a dimensionless phenomenological number and the 
precise form of its magnitude as a function of the inhomogeneities 
present in the lattice depends on the microscopic details.
From Fig.~\ref{fig:pd} it also follows that effect of the disorder 
is offset by spinon flavor numbers, and, therefore, stronger disorder 
is needed to drive the confinement transition. However, more 
generic nonsymmetric random perturbation in the disorder coupling has a
runaway flow to strong coupling (such as in Fig.~\ref{fig:flow})
and, consequently, the Dirac spin-liquid state suffers a smearing transition to 
a symmetry broken phase for any magnitude of the disorder strength.

\section{Conclusion}
\label{conclusion}

From the available studies \cite{thomson17,goswami17,zhao17}
it is known that the most generic 
local random perturbations 
to two-dimensional Dirac spin-liquid Hamiltonians with noncompact 
U(1) gauge fields lead to 
strong disorder instabilities and, consequently, emergence of a confined 
phase. In a realistic material, this outcome can, however, be affected 
by two key aspects,
the symmetric nature of the random perturbations \cite{thomson17}
and the effects of the microscopic monopole excitations of the compact
gauge field fluctuations that arise in a lattice \cite{polyakov77}. 
Our paper extends the findings of the earlier works on this problem
to consider the interaction between quenched disorder and the 
monopole operators. 

We have adapted the radial quantization techniques to 
calculate the scaling dimensions of local primary operators in 
a $(2+1)$-dimensional
CFT with quenched random couplings.
By computing the renormalized scaling dimension
of the monopole operators of the effective $\rm cQED_{2+1}$ description, 
we establish that 
the spin-liquid is further destablized due to the disorder
induced under-screening of the monopole operators. In the absence
of monopoles, Ref.~\cite{thomson17} found that the random vector
chiralitylike microscopic perturbations, which break the 
microscopic SU(2) spin symmetry down to U(1) and introduces 
random current perturbations to the effective theory, flows to a 
finite disorder fixed point
where the spin-liquid may yet survive despite some quantitative
modifications \cite{vafek08}. We find that this finite disorder
fixed point is in fact also fragile once the monopole operators are 
considered. On the other hand, more generic forms of random 
perturbations are seen to drive the RG flow towards 
a strong coupling fixed point where both the monopole fugacity 
and disorder strength turns relevant. Our paper carries an important 
input towards the search of U(1) Dirac spin-liquid phases in frustrated 
two-dimensional spin systems, especially on the triangular lattice 
where it is anticipated that a monopole-driven confinement of the gapless 
spinon excitations is avoided due to the lattice's 
nonbipartite nature \cite{song19,song20}. 

On the experimental side, observations of
spin-liquid like signatures in certain 
triangular lattice organic salts such as 
$\rm \kappa-(ET)_2Cu_2(CN)_3$ \cite{shimizu03,yamashita08}
and 
$\rm EtMe_4Sb[Pd(dmit)_2]_2$ \cite{itou10,yamashita10}
have triggered discussions surrounding the viability of a stable
U(1) Dirac spin-liquid phase at zero temperature. Similarly, 
experimental studies on compounds such as  
Herbertsmithite have inspired the possibility of observing a 
quantum spin-liquid ground state on the kagome lattice 
\cite{norman16}. Although, 
the nature of the non-magnetic ground state of the kagome lattice
Heisenberg antiferromagnet is 
still a matter of active debate \cite{norman16}, some very recent 
studies indicate that the U(1) Dirac spin-liquid state is preferred
over other candidate $\rm Z_2$ spin-liquid states \cite{jiang19}.
However, many of the 
prospective triangular lattice and kagome lattice 
spin-liquid compounds are also noted
to include significant quenched randomness effects 
which may 
mimic spin-liquid behaviors
\cite{shimokawa15} or show prominent spin-glass type signatures \cite{ma18}.  

For time-reversal invariant random exchange like perturbations, the 
effective CFT describing U(1) Dirac spin-liquids has an RG flow to 
strong disorder coupling where the spin-liquid is purportedly 
destabilized \cite{thomson17,zhao17}. Our study indicates that time-reversal
invariant random vector-chirality perturbations generically
turn the symmetry allowed monopole operators relevant on the
triangular and kagome lattices, thereby introducing strong instabilities 
to the spin-liquid and confining the Dirac-dispersing spinons.  
This result unambiguously establishes that in the presence of a 
 large class of generic 
random perturbations, a Dirac spin-liquid is not stable 
in two-dimensional
non-bipartite lattices as the monopoles which are irrelevant in the 
clean limit turn relevant due to disorder effects and confine the 
perturbed theory. 

\section{Discussion and Outlook}
\label{discussion}
 
Now, we turn our attention to the nature of the disordered phase 
that emerges when the monopoles become relevant due to quenched random 
perturbations.
The microscopic monopole operators of the compact U(1) 
Dirac spin-liquid states on two-dimensional 
non-bipartite lattices are either 
spin-singlet or spin-triplet 
excitations \cite{song20}. In a clean frustrated spin system, 
the proliferation of 
the singlet-type monopoles
is associated with VBS ordering (which breaks the lattice symmetries) 
whereas 
the triplet monopole proliferation leads to spiral Ne\'{e}l ordering 
\cite{song19,song20}.
Although 
the complicated interactions among the quenched random perturbation, 
the monopole operators and the order parameters of these competing 
long-range ordered phases are difficult to track, some recently established 
no-go results on disordered frustrated spin systems can 
help us understand the character of the disorder-driven confined phases 
proximate to the 
two-dimensional Dirac spin-liquid phases on such lattices. 

For frustrated SU(2) [or U(1)] symmetric 
spin systems in two spatial dimensions 
(e.g. the triangular lattice Heisenberg model), 
it has been shown that both spiral N\'{e}el and 
VBS orders are unstable against small random
exchange perturbations and ultimately give rise to short or 
quasi-long-ranged ordered glassy phases \cite{kimchi18,dey20,dey20b}. 
Following the work in Ref.~\cite{utesov15}, the same inference 
can be extended to random vector-chiralitylike perturbation 
effects on spiral ordering. 
SU(2) symmetric random exchange couplings, which
are associated with spin-singlet random perturbations to the effective
theory of the Dirac spin-liquid (see Sec.~\ref{sec:disorder_model}),
would naturally lead to the proliferation of 
spin-singlet monopole operators and, consequently, an instability to 
VBS-type ordering in the disordered background. However, 
following the recent arguments on 
frustrated two-dimensional spin 
systems even weak disorder leads to the destruction of long-range 
VBS ordering in favor of domain formation and nucleation of spinons 
\cite{kimchi18}. For intermediate to strong disorder a
glassy random singlet ground state has been observed in numerical 
simulations of the random-bond triangular lattice Heisenberg antiferromagnet
\cite{wu19}. On the other hand, random vector-chirality 
like couplings, which generate disordered spin-triplet perturbations 
(Sec.~\ref{sec:disorder_model}), lead to proliferation of
spin-triplet monopole operators and, consequently, 
introduce a putative magnetic spiral ordering instability. 
However, based on the recent results, such spiral 
ordering on frustrated spin systems are actually 
destabilized in favor of 
a spin-glass phase 
for weak exchange disorder \cite{dey20}.

Altogether, it is, therefore plausible, that in the presence of 
generic random perturbations, the Dirac spin-liquid 
ground states of the triangular and kagome lattice Heisenberg 
antiferromagnets
are destabilized, most likely in favor of short-range ordered 
ground states 
where monopole operators are confining and disorder flows
to strong coupling. This observation is 
consistent with the quantum critical behavior seen in the
compound $\rm \kappa-(ET)_2Cu_2(CN)_3$ where it is anticipated
that a gapless spin-liquid state enters a glassy phase in the
presence of random Dzyaloshinskii-Morya and 
multi-spin chiral interaction 
at low temperatures \cite{riedl19}. However, a more 
accurate characterization 
of the disordered phases proximate to the Dirac spin-liquid
phases on these non-bipartite lattices 
requires further microscopic studies which have been left as 
future tasks.

Finally, it is to be noted that the methodology discussed in this 
treatment have wider applicabilities. 
The radial quantization scheme with quenched random coupling  
can be adapted to a number of $(2+1)$-dimensional 
U(1) conformal gauge theories \cite{pufu13} perturbed by quenched
disorder. Among them, the $\mathbb{CP}^{N_b-1}$ theory of 
unit-norm $N_b$-component complex bosonic spinons constitute a 
viable example. This theory captures the transition between 
collinear N\'eel ordering and VBS ordering on bipartite 
quantum antiferromagnets \cite{dyer15}.
The monopole operators of this model are
interpreted as the order parameter of VBS order. 
Therefore, it will be worthwhile to apply and extend the methodology of 
this paper 
within the context of the collinear N\'{e}el to 
VBS transition in the 
presence of quenched disorder and investigate how the role of the 
monopole operators are affected at the critical point.

\begin{acknowledgments}
    We gratefully appreciate the instructive discussions 
    with A. Kapustin, D. Bernard,  
    and S. M. Chester on related topics. The very helpful communications 
    with M. Mezei, \'{E}. Dupuis and W. Witczak-Krempa during the preparation
    of the paper are particularly acknowledged. A special thanks 
    goes to M. Vojta for the valuable
    suggestions along the way which significantly shaped the focus of 
    this paper. 
    We acknowledge financial support from the Deutsche Forschungsgemeinschaft
    through SFB 1143
    (Project No. 247310070) and the W\"urzburg-Dresden Cluster of
    Excellence on Complexity and Topology in Quantum Matter-
    \textit{ct.qmat} (EXC 2147, Project No. 39085490).
\end{acknowledgments}

\appendix

\section{Regularization of the double summation in the disordered $\rm QED_{2+1}$ free energy}
\label{app:double_sum}

The second contribution appearing in the expression for the free
energy Eq.~\eqref{eq:free_energy} involves a
formally divergent
double summation 
over angular momentum indices. In this
appendix, the divergent summation is regularized using 
the $\zeta$
function method. 
It is convenient to split the
double summation in two parts,
\begin{equation}
\begin{aligned}
I_2(q)&=\sum_{l,l'=q+1}^\infty\frac{ll'}
{\sqrt{l^2-q^2}+\sqrt{{l'}^2-q^2}}\\
&=\sum_{l=q+1}^\infty
\frac{l^2}{2\sqrt{l^2-q^2}}
+\sum_{\substack{l,l'=q+1
		\\l\neq l'}}^\infty
\frac{ll'}{\sqrt{l^2-q^2}+\sqrt{{l'}^2-q^2}}.
\end{aligned}
\end{equation}
The first single summation grows as $\propto l$ asymptotically. Following
the same regularization technique as in
the main text a perfectly converged summation may
instead be considered,
$(1/2)\sum_{l=q+1}^\infty l^2/(l^2-q^2)^{1/2+s}$.
It is possible to subtract and then add back from this expression its 
asymptotic dependence and then analytically continue the result 
to $s\rightarrow 0$ using the 
identities of the Hurwitz $\zeta$ function. 

However, in certain cases where
the resulting expression contains essential singularities in the limit $
s\rightarrow 0$, a modification to this regularization scheme is 
better suited \cite{chester16}. Let us consider $A(s)$ as a quantity
which we want to analytically continue to $\lim_{s\rightarrow 0}A(s)=a_0$ 
but $A(s)=a_{-m}s^{-m}+a_{-(m-1)}s^{-(m-1)}+\dots a_0+a_1 s +\dots$ is 
singular. We can instead take the operator, 
\beq
\mathcal{D}\left[\frac{d^n}{ds^n}\left(s^n A(s)\right)\right]
=a_0,
\eeq  
and consider any $n> m$ such that the regularized finite part 
$a_0=\lim_{s\rightarrow 0}A(s)$ is obtained without encountering any 
singularities. The original scheme corresponds to $n=0$.

Now following this strategy the manipulated summand yields the regularized finite
contribution
\begin{equation}
\begin{aligned}
&\frac{1}{2}\sum_{l=q+1}^\infty
\left(
\frac{l^2}{(l^2-q^2)^{1/2+s}}
-\left[l^{1-2s}
+\left(s+\frac{1}{2}\right)\frac{q^2}{2}l^{-1-2s}
\right]
\right)_{s\rightarrow 0}\\
&+\frac{1}{2}
\sum_{l=q+1}^\infty
\left[l^{1-2s}
+\left(s+\frac{1}{2}\right)\frac{q^2}{2}l^{-1-2s}
\right]_{s\rightarrow 0},\\
=&
R_2(q)
+
\frac{1}{2}
\left[
\zeta(-1,q+1)-\frac{q^2}{2}\psi(q+1)
\right].
\end{aligned}
\end{equation}
Here $R_2(q)=\frac{1}{2}\sum_{l=q+1}^\infty
\left(\frac{l^2}{\sqrt{l^2-q^2}}
-\left[l+\frac{q^2}{2l}\right]\right)
$ is now a convergent summation even after taking the
limit $s\rightarrow 1$.

The remaining double summation offers more difficulty.
To make progress the summation may be cast in a different form
\beq
\sum_{l\neq l'}\frac{ll'\left(\sqrt{l^2-q^2}
	-\sqrt{{l'}^2-q^2}
	\right)}{l^2-{l'}^2}
&=\sum_{l\neq l'}\frac{2ll'}{l^2-{l'}^2}\sqrt{l^2-q^2},
\eeq
which is helpful to obtain the summation over one of the indices
in a purely analytical form. Thus, the summation over $l'$ of the quantity 
$\frac{2l'}{l^2-{l'}^2}$ is first considered. The summand grows as
$\propto -\frac{2}{l'}$ and a finite value to it can be assigned 
by considering the modified regularization $\mathcal{D[\cdots]}$. 
For $l\geq q+2$ it follows that,
\begin{equation}
\begin{aligned}
\sum^\infty_{\substack{
		l'=q+1
		\\
		l'\neq l}}
\frac{2l'}{l^2-{l'}^2}
&=\sum^\infty_{\substack{
		l'=q+1
		\\
		l'\neq l}}
-\left(\frac{1}{l'-l}+\frac{1}{l'+l}\right),
\\
&=\frac{1}{2l}+\psi(l-q)+\psi(l+q+1)
\ \forall \ l \geq q+2, 
\end{aligned}
\end{equation}
which then reduces the double summation to a single summation over $l$,
\begin{equation}
\begin{aligned}
&\sum^\infty_{\substack{l,l'=q+1\\l\neq l'}}
\frac{ll'}{\sqrt{l^2-q^2}+\sqrt{{l'}^2-q^2}}
\\
=&
\sum_{l'=q+2}^\infty
\frac{2(q+1)l'\sqrt{2q+1}}{(q+1)^2-{l'}^2}
+\sum_{\substack{
		l=q+2,l'=q+1\\
		l'\neq l}}^\infty
\frac{2ll'}{l^2-{l'}^2}\sqrt{l^2-q^2}
\\
=&
\sum_{l=q+2}^\infty
\frac{2l(q+1)\sqrt{2q+1}}{(q+1)^2-l^2}\\
&+
\sum_{l=q+2}^\infty
l\sqrt{l^2-q^2}
\Big(
\frac{1}{2l}
+\psi(l-q)
+\psi(l+q+1)
\Big).
\end{aligned}
\end{equation}
The first term in the above expression can be computed similarly as
a principle value to yield the finite contribution,
\beq
&-(q+1)\sqrt{2q+1}
\sum_{l=q+2}^\infty
\Big(
\frac{1}{l-(q+1)}
+
\frac{1}{l+(q+1)}
\Big)\\
&=(q+1)\sqrt{2q+1}
\left[\gamma+\psi(2q+3)\right]
\eeq
where $\gamma=0.57721\dots$ is the Euler-Mascheroni constant.

The second term is also formally divergent due to its asymptotic growth,
$\propto 2 l^2 \ln l+l/2 -q^2\ln l -(1/6+q+q^2) -q^2/(4l)  $.
The logarithmically growing portion can be regularized by using the
identity, $\ln l =-\frac{d }{ds}l^{-s}|_{s=0}$, such that one has,
\begin{equation}
\begin{aligned}
\sum_{l=q+2}^\infty \ln l&=
-\frac{d}{ds}\zeta(s,q+2)|_{s=0}
=-\zeta'(0,q+2).
\end{aligned}
\end{equation}
In a similar manner the other logarithmically growing term gets the
finite expression,
$\sum_{l=q+2}^\infty l^2\ln l=-\zeta'(-2,q+2)$. The terms which
grows as $l$ and $1/l$ can be regularized using the various
identities already invoked above. Subtracting the diverging part
from the summand and adding its regularized value back to the summation
as above, the regularized double summation is thus obtained to be
\begin{equation}
\begin{aligned}
&\sum^\infty_{\substack{l,l'=q+1
		\\
		l\neq l'
}}\frac{ll'}{\sqrt{l^2-q^2}+\sqrt{{l'}^2-q^2}}\\
=&
(q+1)\sqrt{2q+1}
\left[\gamma+\psi(2q+3)\right]
+R_3(q)\\
&+\Big[-2\zeta'(-2,q+2)
+\frac{\zeta(-1,q+2)}{2}
+q^2\zeta'(0,q+2)\\
&-\left(\frac{1}{6}+q+q^2\right)\zeta(0,q+2)
+\frac{q^2}{4}\psi(q+2)
\Big],
\end{aligned}
\end{equation}
where
\begin{equation}
\begin{aligned}
R_3(q)=& \sum_{l=q+2}^\infty
\left(l\sqrt{l^2-q^2}
\left[
\frac{1}{2l}
+\psi(l-q)
+\psi(l+q+1)
\right] \right.\\
&\left.-\left[
2l^2\ln l +\frac{l}{2}
-q^2\ln l
-(1/6+q+q^2)
-\frac{q^2}{4l}
\right]
\right)
\end{aligned}
\end{equation}
is once again a convergent sum. Putting together
all of these pieces the complete and finite regularized
expression for the second contribution to the disorder-averaged scaling
dimension [Eq.~\eqref{eq:scaling_dimension}] is found to be
\begin{equation}
\begin{aligned}
I_2(q)-I_2(0)
&=R_2(q)
+\left[R_3(q)-R_3(0)\right]
+\left[f(q)-f(0)\right],
\end{aligned}
\end{equation}
where $f(q)$ combines the contributions with an analytical
expression,
\begin{equation}
\begin{aligned}
f(q)=&(q+1)\sqrt{2q+1}H_{2q+2}\\
&+\frac{q^2}{2}\ln\frac{\Gamma(2+q)^2}{2\pi}\\
&-\frac{4-q\left(4+35q+12\left(3+q\right)q^2\right)}
{12(1+q)}\\
&-2\zeta'(-2,q+2).
\end{aligned}
\end{equation}
Here $H_z$ is the harmonic number and $\Gamma(z)$ is the $\Gamma$ function,
and both of these special functions can be evaluated up to arbitrary
precision.

\bibliography{disu1sl}

\end{document}